\documentclass{sf2a-conf}
\usepackage{graphicx}
\usepackage{natbib}

\begin{document}

\TitreGlobal{SF2A 2009}

%%-----------------------------
%%      the top matter
%%-----------------------------
\title{The Gaia mission and variable stars}
\author{Eyer, L.}\address{Geneva Observatory, University of Geneva, 1290 Sauverny, Switzerland}
\author{Mowlavi, N.}\address{ISDC, Geneva Observatory, University of Geneva, 1290 Versoix, Switzerland}
\author{Varadi, M.$^1$}
\author{Spano, M.$^1$}
\author{Lecoeur-Taibi, I.$^2$}
\author{Clementini, G.}\address{INAF Osservatorio Astronomico di Bologna,  40127 Bologna, Italy }
\runningtitle{The Gaia mission and the variable stars}
\setcounter{page}{237}
% Keep this line, even if the page will be settled afterwards..

\index{Eyer, L.}
\index{Mowlavi, N.}
\index{Varadi, M.}
\index{Spano, M.}
\index{Clementini, G.}

% Repeat the authors here, this will help to make the final index

\maketitle

\begin{abstract} 
The Gaia satellite, to be launched in 2012, will offer an unprecedented survey of the whole sky down to magnitude 20. The multi-epoch nature of the mission provides a unique opportunity to study variable sources with their astrometric, photometric, spectro-photometric and radial velocity measurements. Many tens of millions of classical variable objects are expected to be detected, mostly stars but also QSOs and asteroids. The high number of objects observed by Gaia will enable statistical studies of populations of variable sources and of their properties. But Gaia will also allow the study of individual objects to some depth depending on their variability types, and the identification of  potentially interesting candidates that would benefit from further ground based observations by the scientific community. 
Within the Gaia Data Processing and Analysis Consortium (DPAC), which is subdivided into 9 Coordination Units (CU), one (CU7) is dedicated to the variability analysis.
Its goal is to provide information on variable sources for the Gaia intermediate and final catalogue releases.
 \end{abstract}

\section{Introduction}
Each object will be observed by Gaia a mean of 70 times during the 5 year mission. For each transit, Gaia will have quasi-simultaneous broad-band (G) photometry, blue (BP) and red (RP) spectro-photometry, and radial velocity spectrometer (RVS) measurements (in half of the cases for this latter instrument). As the shortest integration time is 4.4 seconds, variable sources can be detected on time scales from tens of seconds to years. The photometric precision should reach the milli-magnitude level at the bright end, and about 20 mmag at a magnitude of 20. In addition, the highly accurate astrometry will provide parallaxes and proper motions that will complement the photometric and RVS data. Most of the known variability types will benefit from the Gaia mission, thanks to its multi-epoch observations. In order to give to the scientific community the opportunity to perform follow-up ground based observations, the Gaia consortium puts in place a system of alerts and intermediate releases. For some events that occur uniquely and on a short time scale, a flux-based alert will be issued by the DAPC Coordination Unit 5 dedicated to the Photometric reduction. Variability announcements that are less time critical, for example those providing a list of candidates of interesting variable sources such as RR Lyrae stars, will be prepared by the Coordination Unit 7 (CU7) responsible of the analysis of all types of variables outside the solar system.

\section{The Gaia scanning law}

Gaia is a survey mission and is scanning the whole sky according to a prescribed law, designed to optimise the astrometric results. The Gaia sampling has been previously described in Eyer \& Mignard (2005). Since that study, some modifications have been brought to the satellite design and to the scanning law. However, the general conclusions for the sampling properties do not change for the astrometric field. 

The general behaviour of the time sampling pattern results from the design and operating mode of the satellite: Gaia has two fields of view separated by 106.5 degrees, and rotates around itself with a period of 6 hours. As a result, a sequence of measurements consists of several transits separated successively by 1h46m and 4h14m, which correspond to the times elapsed from one field to the other. The next sequence of transits appears about 1 month later, due to  the rotation axis precession and the satellite  orbital motion. Between 40 and 250 per transit measurements will thus be collected  for each star during the five year mission, depending on its ecliptic latitude, with a predicted mean number of 80 measurements. If we take into account "dead times", a recent study shows that the expected mean number of measurements lowers to 70 (de Bruijne 2009). 

\begin{figure}[h]
   \centering
   \includegraphics[width=17cm]{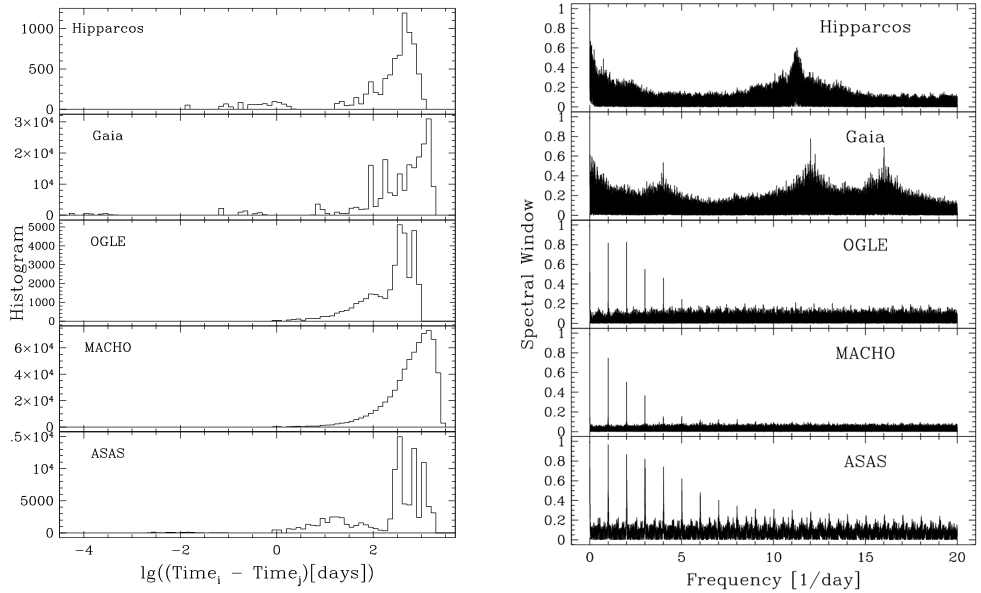}
      \caption{Sampling properties of different missions and projects. Left: Histograms of the time differences between two successive measurements for a randomly chosen star, per mission or project. Right: Spectral windows of different surveys. The predicted time lags for Gaia are given based on per CCD photometry.}
       \label{TimeLag_SpectWindow}
   \end{figure}

The Gaia time sampling is very similar to the time sampling of Hipparcos since their scanning laws were built on the same principle, but it significantly differs from the time sampling of ground based photometric surveys. In Fig.~\ref{TimeLag_SpectWindow}, we present the sampling properties of different missions and projects for a randomly chosen star. The spectral window of Gaia varies quite a lot from one region of the sky to another. We also remark that the high amplitude peaks in the spectral window, which are causing aliases in the Fourier space, are located at high frequencies for Gaia, as it is for Hipparcos.
  
\section{Periodic and short periods variable stars }
The detection rate of mono-periodic signals observed by Gaia is expected to be quite good for a wide range of periods. Eyer \& Mignard (2005) showed that this period recovery for regular variable stars depends on the ecliptic latitude, reaching more than 95\% over more than 40 degrees of ecliptic latitudes and for S/N ratios as low as 1.3.

The Gaia time sampling and the CCD data acquisition scheme allow in principle to probe stellar variability also on time scales as short as several tens of seconds, thereby giving potential access to the study of short-period (less than a few hours) variable stars in a large and homogenous sample of stars. In order to explore that time scale regime, Varadi et al. (2009), in a first step, extended the work of Eyer \& Mignard (2005) to periods shorter than two hours, and showed that the period recovery of a sinusoidal signal with a Gaia time sampling is above 90\% for S/N ratios as low as 1.0, provided that per-CCD photometry is used. A second step has been initiated by Mary et al. (2006) to introduce multi-periodic sinusoidal signal, simulating the case of the roAp star HR 3831. Simulating 16 frequencies for that star, they were able to recover three frequencies from a noiseless curve. The third step consists in testing the recovery capability of non-linear multi-periodic light curves. The study is performed on simulated  light curves of ZZ Ceti stars (Varadi et al. 2009) and takes into account the flux transfer of a sinusoidal signal from the base of the convective envelope of those stars to their surface. The results of those simulations show that the non-linear effect introduced by the flux transfer through the envelope degrades by only a few percents the performance of the recovery rate of the main period for multi-periodic ZZ Ceti stars. The next step should consider the case of non-stationarity of variability that characterises several classes of short pulsators. In these cases, the stellar pulsation periods and amplitudes can change on time scales from weeks to years. Further studies are under way to analyse the impact of those effects on the Gaia detection capability of those stars.

\section{Pseudo-periodic and irregular variable objects}
Due to the nature of the Gaia sampling, the behaviour of irregular and pseudo-periodic variable objects poses many challenges. First, their irregular nature makes their characterisation particularly difficult. Some methods such as the structure function/variogram (Eyer \& Genton 1999) can help characterising the variability time-scales present in the source. This technique was applied for example by Eyer (2002) to search QSOs in OGLE-II database. Second, they may "contaminate" the sample of periodic variable stars. The analysis of the pseudo-periodic stars can indeed identify spurious frequencies from their Fourier spectrum and wrongly classify them as periodic variables. This lowers the quality of the catalogue of periodic variable objects.

The analysis of irregular or pseudo-periodic variable objects is however interesting, as it can lead to the detection of rare cases of variable objects. An example is given by the secular variable stars such as post-Asymptotic Giant Branch stars.These stars are evolving so fast that the photometric variations due to their stellar evolution  can become detectable on human time scales. Few such stars have been seen to cross the entire colour-magnitude diagram in some decades. In Gaia, a work package is dedicated to the detection and characterisation of such stars based on the search of global changes in their magnitude or colour. Preliminary  studies are carried out in existing surveys such as OGLE (Spano et al. 2009) and EROS.

\section{Transients}
The detection of transient events is also challenging. About 6,000 supernovae, for example, are expected to be detected by Gaia down to magnitude 19 (Gilmore 2009), with one third of them being detected before maximum light. While they are not likely to be a source of contamination for the catalogue of periodic stars, their possible confusion with other types of non-periodic stars remains to be addressed, and an adequate procedure should be put in place for their detection. 
%However, if no data is acquired by the mission during the transient event, the stars will not be problematic contaminants for the periodic objects. Obviously here also there is a difference between the signal presence and its correct detection. A procedure should be established to see confusion matrices between supernovae and main contaminants.

%\subsection{An example of transients: microlensing events}
Microlensing events are other transient phenomena that are of potential interest for Gaia. Over the 1988 microlensing events detected in OGLE-III, 66\% (1324 events) have at least one measurement within the lensing event (Wyrzykowski 2009). For events with time-scales longer than 
30 days the statistics improves to 93\%.
The automatic detection and fast identification of microlensing events are not obvious though, despite the fact that they have clear  signatures with a smooth and achromatic rise and fall. An algorithm is being set up to detect such events (Eyer et al. 2009). A preliminary comparison of our microlensing event candidates with those of Wozniak (2001) indicates a high recovery rate on OGLE-II data.
The application of our algorithm to the Hipparcos catalogue resulted in only few false detections, showing that the robustness of the identification procedure.

\section{Variable stars simulations}
In order to test the algorithms that are set up in CU7 to detect and characterise the variable objects observed by Gaia, simulated light curves are produced for an increasing number of types of variable stars (Mowlavi  2009). Currently, Cepheids, RR Lyrae of types ab and c, delta Scuti stars, ACVn stars, Miras, roAp stars, semi-regular variables, ZZ Ceti stars, dwarf novae, active galactic nuclei and microlensing events are simulated.

The simulated light curves, together with the properties of each type of variable stars (their location in the HR diagram and the probability of their occurrence), are provided to Coordination Unit 2 (CU2) in order to feed their Gaia simulation code.
Simulated Gaia time series of variable stars, as realistic as possible, are thus aimed to test the CU7 software.

\section{Ground based Observations}
The Gaia DPAC may need some ground based data to help the preparation of its data 
processing. A Working Group, Ground-Based Observations for Gaia (GBOG), has been 
formed and is establishing the need of such observations and is also coordinating 
the proposals to ESO. For the variability analysis, it has been felt that there 
is a wealth of data which are already available, e.g. Hipparcos, MACHO, EROS, 
OGLE, CoRoT, HAT, SDSS data and therefore there is no need to gather additional 
data for the moment.

A network of 12 telescopes of 1-2 m size is currently in place within CU7. It 
is worth mentioning that the use of such 1-2 m class telescopes is particularly 
adequate also for the photometric follow-up studies of variable stars. Within 
DPAC, such follow-up should be only done for validation purpose. However these 
telescopes could be used for the scientific exploitation, once the data is public. 
Spectroscopic studies of bright variables such as Cepheids, Long Period variables 
or RR Lyrae stars will also benefit from those small size telescopes.

\begin{acknowledgements}
We would like to thank the whole Gaia team in Geneva:
M.~Beck,   P.~Dubath,        L.~Guy,
A.~Jan,    K.Nienartowicz, L.~Rimoldini,
as well as the CU7 members who are contributing to the work on the CU7 Variability Processing and Analysis. We also want to thank J.~de~Bruijne for discussions on the scanning law.
\end{acknowledgements}

%Bohr, N., Einstein, A., & Fermi, E. 1992, MNRAS, 301, 257 (BEF)
% Curie, M., & Curie, P. 1991, A&A, 248, 612
% de Gaulle, C. 1996, Solar Phys. (Oxford: Oxford Univ. Press)
% Heisenberg, W., & West, C. N. 1993, Australian J. Phys., 537, 36  (Paper III)
% Laurel, S., & Hardy, O. 1994, Active Galactic Nuclei, in The Evolution
% and Distribution of Galaxies, ed. W. Churchill, F. D. Roosevelt, & J.
% Stalin (New York: Wiley), 210


\begin{thebibliography}{}
\bibitem{} de Bruijne 2009, private communication
\bibitem{} Eyer L., \& Mignard F. 2005, MNRAS 361, 1136
\bibitem{} Eyer L., Evans D.W., \& Lecoeur-Taibi I. 2009, Gaia document: GAIA-C7-SP-GEN-LE-008
\bibitem{} Eyer L., \& Genton M.G. 1999, A\&AS 136, 421
\bibitem{} Eyer L. 2002, AcA 52, 241
\bibitem{} Gilmore G. 2009, March Kick-off GREAT meeting, Cambridge, UK
\bibitem{} Holl B., Hobbs D., \& Lindegren L. 2009, IAU 261, 1703
\bibitem{} Mary D. et al. 2006, Gaia document: GAIA-C7-TN-ARI-DM-003-1
\bibitem{} Mowlavi N. 2009, Gaia document: GAIA-C7-UG-GEN-NM-006
\bibitem{} Spano M., Mowlavi N., Eyer L., \& Burki G. 2009, in Stellar Pulsation, Challenges for Theory and Observation, AIP, in press 
\bibitem{} Varadi M., Eyer L., Jordan S., Mowlavi N., \& Koester D. 2009, in Stellar Pulsation, Challenges for Theory and Observation, AIP, in press
\bibitem{} Wozniak P.R., Udalski A., Szymanski M., et al. 2001, Acta Astronomica, 51, 175
\bibitem{} Wyrzykowski L. 2009, Gaia document: GAIA-C5-TN-IOA-LW-001


\end{thebibliography}
\end{document}